\documentclass[11pt, twocolumn]{aastex631}
\usepackage{CJK}
\usepackage{bm}
\usepackage{units}

\usepackage{float}

\usepackage{pgfplots}
\usepackage{multirow}
\pgfplotsset{compat=1.15}
\newcommand\bluesout{\bgroup\markoverwith{\textcolor{blue}{\rule[0.5ex]{2pt}{0.4pt}}}\ULon}

\date{\today}

\begin{document}
\begin{CJK*}{UTF8}{gbsn}

\title{$1/f$ Noise in the Heliosphere: A Target for PUNCH Science}
\author[0009-0008-8723-610X]{Jiaming Wang (王嘉明)}
\affiliation{Department of Physics and Astronomy, University of Delaware}
\author[0000-0001-7224-6024]{William H. Matthaeus}
\affiliation{Department of Physics and Astronomy, University of Delaware}
\author[0000-0002-7174-6948]{Rohit Chhiber}
\affiliation{Department of Physics and Astronomy, University of Delaware}
\affiliation{Heliophysics Science Division, NASA Goddard Space Flight Center}
\author[0000-0003-3891-5495]{Sohom Roy}
\affiliation{Department of Physics and Astronomy, University of Delaware}
\author[0009-0005-9366-6163]{Rayta A. Pradata}
\affiliation{Department of Physics and Astronomy, University of Delaware}
\author[0000-0003-4168-590X]{Francesco Pecora}
\affiliation{Department of Physics and Astronomy, University of Delaware}
\author[0000-0003-2965-7906]{Yan Yang}
\affiliation{Department of Physics and Astronomy, University of Delaware}


\begin{abstract}
We present a broad review of $1/f$ noise observations in the heliosphere, and discuss and complement the theoretical background of generic $1/f$ models as relevant to NASA's Polarimeter to UNify the Corona and Heliosphere (PUNCH) mission. First observed in the voltage fluctuations of vacuum tubes, the scale-invariant $1/f$ spectrum has since been identified across a wide array of natural and artificial systems, including heart rate fluctuations and loudness patterns in musical compositions. In the solar wind, the interplanetary magnetic field trace spectrum exhibits $1/f$ scaling within the frequency range from around $\unit[2 \times 10^{-6}]{Hz}$ to around $\unit[10^{-3}]{{Hz}}$ at 1 au. One compelling mechanism for the generation of $1/f$ noise is the superposition principle, where a composite $1/f$ spectrum arises from the superposition of a collection of individual power-law spectra characterized by a scale-invariant distribution of correlation times. In the context of the solar wind, such a superposition could originate from scale-invariant reconnection processes in the corona. Further observations have detected $1/f$ signatures in the photosphere and corona at frequency ranges compatible with those observed at 1 au, suggesting an even lower altitude origin of $1/f$ spectrum in the solar dynamo itself. This hypothesis is bolstered by dynamo experiments and simulations that indicate inverse cascade activities, which can be linked to successive flux tube reconnections beneath the corona, and are known to generate $1/f$ noise possibly through nonlocal interactions at the largest scales. Conversely, models positing in situ generation of $1/f$ signals face causality issues in explaining the low-frequency portion of the $1/f$ spectrum. Understanding $1/f$ noise in the solar wind may inform central problems in heliospheric physics, such as the solar dynamo, coronal heating, the origin of the solar wind, and the nature of interplanetary turbulence.

\end{abstract}

\section{Introduction}

$1/f$ noise, otherwise known as ``flicker noise'', refers to a signal in which the amplitude of the spectral density $P(f)$ inversely scales with the frequency $f$. Such a spectrum has the unique property that the integrated power per octave remains constant across frequencies. Mathematically, the integration of the spectrum over a frequency range $f_1$ to $f_2$, i.e. $\int P(f) df \sim \int df/f \sim \log{f_2/f_1}$, depends solely on the ratio $f_2/f_1$. This total power is insensitive to rescaling of the frequency by an arbitrary factor, a reflection of the {\it scale invariance} of the power $df/f$. The distribution $P(f) \sim 1/f$ is often referred to as a scale-invariant distribution and is observed across a wide array of physical systems~\citep[see, e.g., review by][]{Dutta81}, including heliospheric plasmas such as the interplanetary magnetic field~\citep{Bavassano82, Burlaga84, Matthaeus86} and elsewhere~\citep{Matthaeus07, Bemporad08}. When the scale-invariant $1/f$ noise spectrum is observed, it is often fruitful to search for a scale-invariant physical process that produces the observation. In this way the study of $1/f$ noise can lead to new physical insights. Some examples of such efforts are given in Section~\ref{sec:2}. 

The primitive observations of $1/f$-like solar wind magnetic field spectra~\citep{Coleman68, Bavassano82, Denskat82} revealed a tendency toward spectral indices shallower than the Kolmogorov value of $-5/3$ at lower frequencies, typically below around a few times $\unit[10^{-3}]{Hz}$ to a few times $\unit[10^{-5}]{Hz}$. However, the significance of $1/f$ spectrum was not fully recognized in these early reports. It was \citet{Burlaga84} who first placed an emphasis on the form $f^{-1}$ and in addition employed data records long enough to probe frequencies down to $\unit[10^{-6}]{Hz}$ and lower. The need for extended data records is a recurrent theme in identifying $1/f$ power spectrum, as discussed further in the following sections.

Understanding the origin of the interplanetary $1/f$ observations was counted among the scientific motivations for design of the Parker Solar Probe (PSP) mission~\citep{Fox16}. As will be discussed in Section~\ref{sec:4}, the interplanetary structures associated with the $1/f$ observations are necessarily very large -- relevant $1/f$ spectrum begins to emerge at around $\unit[2 \times 10^{-6}]{Hz}$ and extends to around $\unit[10^{-3}]{Hz}$, corresponding to about the typical reciprocal spacecraft-frame correlation time at 1 au~\citep{Matthaeus82, Isaacs15} and spanning up to more than the solar wind transit time to 1 au.

NASA's Polarimeter to UNify the Corona and Heliosphere (PUNCH) mission will likely, in principle, contain information relevant to the nature of the $1/f$ noise observed {\it in situ}. But will PUNCH be able to detect and characterize its source? At present, this is unclear, as the methods for unambiguously translating the PUNCH imagery into spectral information are still subjects of ongoing research~\citep[see][]{Pecora24}. In anticipation of such developments, we point out this opportunity to use PUNCH to understand at some level the origin of the interplanetary $1/f$ noise. In this paper, we review a history of ideas for how $1/f$ emerges generically across various physical processes (Section~\ref{sec:3}) and make some existing and potential connections to solar wind and coronal observations. Our aim is to provide a foundational background for researchers who will mine PUNCH data for evidence concerning the origins of interplanetary $1/f$ noise.

\section{Background and Examples}
\label{sec:2}

\begin{figure}
\centering\includegraphics[angle=0,width=\columnwidth]{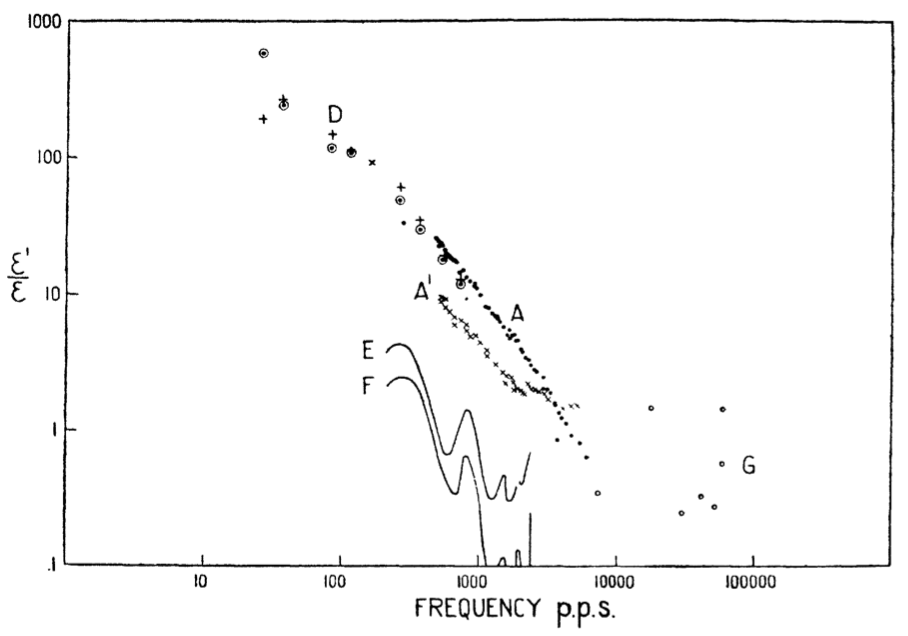}
    \caption{Spectrum of voltage fluctuations in vacuum tubes, exhibiting a range of $1/f$ scaling~\citep[][Fig.~6]{Johnson25}.}
\label{fig:Johnson}
\end{figure}

Low-frequency $1/f$ spectrum was first observed by \citet{Johnson25} when studying voltage fluctuation noise in vacuum tubes. The spectrum, such as the one shown in Fig.~\ref{fig:Johnson}, was subsequently analyzed by \citet{Schottky26} and termed ``flicker noise''. The spectrum was postulated to originate from events of foreign molecules incident upon and disrupting the electron-emitting cathodes, consequently generating voltage fluctuations with Lorentzian spectral profiles. Superposition of these Lorentzian profiles were later shown in many works~\citep[see, e.g.,][]{Bernamont37, vandeZiel50, Machlup81} to lead to $1/f$ spectral behavior.

\begin{figure}
\centering\includegraphics[angle=0,width=\columnwidth]{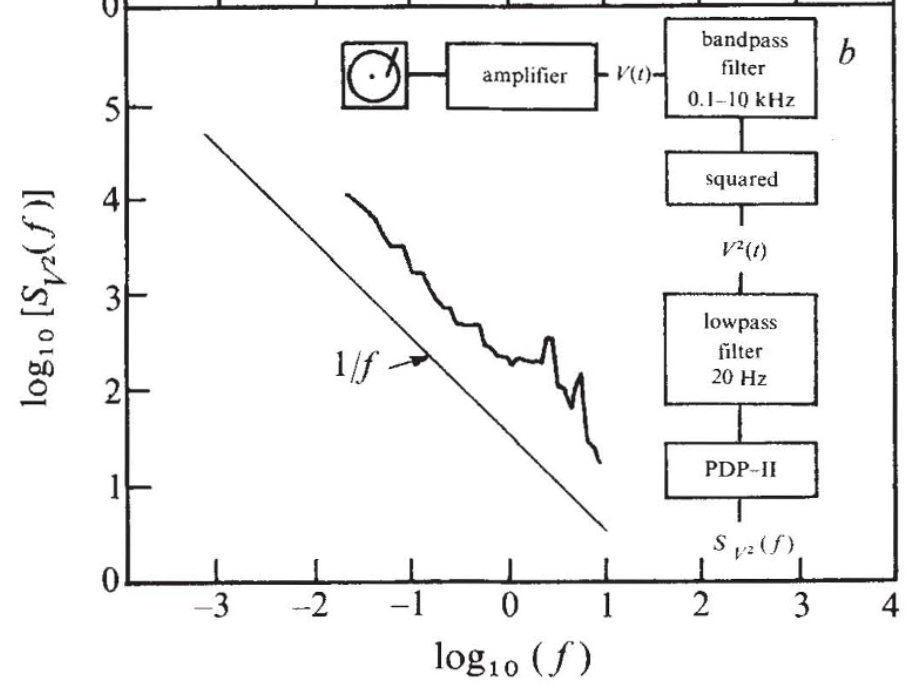}
    \caption{Spectrum of loudness fluctuations of Bach's 1st Brandenburg Concerto~\citep[][Fig.~1]{Voss75},
    consistent with a 
    $1/f$ spectrum, as indicated.}
\label{fig:Voss}
\end{figure}

\begin{figure}
\centering\includegraphics[angle=0,width=\columnwidth]{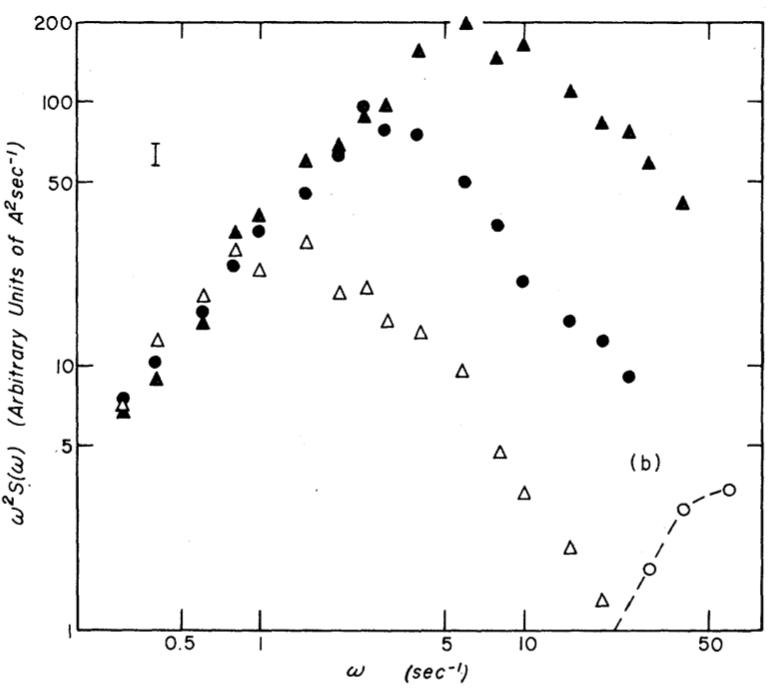}
    \caption{Spectra of noise voltage in turbulent He II exhibiting $1/\omega$ scaling below some critical frequencies, reciprocal of system's relaxation times~\citep[][Fig.~4]{Hoch75}.}
\label{fig:Hoch}
\end{figure}

\begin{figure}
\centering\includegraphics[angle=0,width=\columnwidth]{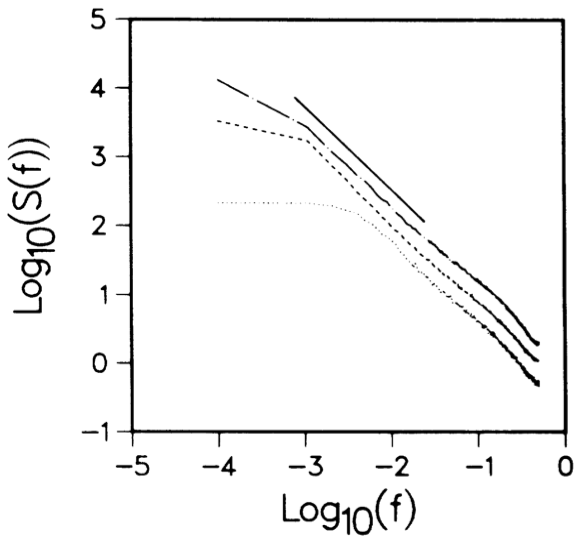}
    \caption{Spectra of total number of particles on simulated two-dimensional lattice models following simplified diffusive dynamics, exhibiting a range of $1/f$ scaling as guided by the solid line~\citep[][Fig.~2]{Jensen90}.}
\label{fig:Jensen}
\end{figure}

Since then, $1/f$ noise has been identified and analyzed in numerous systems, including semiconductors~\citep{vandeZiel50, Caloyannides74}, music and speech~\citep{Voss75, Levitin12}, human heartbeats and cognition~\citep{Musha82, Gilden95}, and more. For instance, the distinctive $1/f$ spectrum is observed in loudness fluctuations and pitch fluctuations, such as in Bach's 1st Brandenburg Concerto~\citep[Fig.~\ref{fig:Voss},][]{Voss75}. Other diverse examples include fluctuations in turbulent He II~\citep[Fig.~\ref{fig:Hoch},][]{Hoch75}, and in cellular automaton and other systems exhibiting self organized criticality~\citep[Fig.~\ref{fig:Jensen},][]{Jensen90}.

In the solar wind, $1/f$ spectrum is often observed to span several orders of magnitude in frequency near 1 au, extending from approximately the reciprocal of the transit time to 1 au (around 100 hours) to more than two decades higher in frequency, around the reciprocal of the correlation timescale~\citep[see, e.g.,][]{Matthaeus82, Matthaeus86, Bruno13}. Beyond this point, the spectrum gradually transitions to the more negative ``$-5/3$'' or ``$-3/2$'' power-law indices associated with the classical descriptions of the inertial range of plasma turbulence~\citep[see, e.g.,][]{Batchelor70, Iroshnikov64, Kraichnan65}. These classical power-law spectra are theoretically grounded in the principle of scale invariance of energy flux across scales within the inertial range, which is defined as a range of scales much smaller than a (fixed) correlation scale and much larger than a typical scale where dissipation comes into play.

One of the arguments for the generation of the $1/f$ part of the spectrum relies on a generic {\it superposition principle}. The principle posits that an ensemble of the so-called ``purely'' random signals with a scale-invariant distribution of correlation times collectively exhibits a $1/f$ spectrum~\citep[][see Section~\ref{sec:3} for details]{Machlup81}. Therefore the presence of $1/f$ signals in the solar wind can be attributed to a scenario in which the correlation scale itself is distributed in a scale-invariant fashion over a sufficiently large range of values. The possibility of a distribution of correlation scales, as opposed to a single scale, becomes relevant for complex systems, apparently including the solar wind~\citep{Ruiz14}, in which the observed fluctuations emerge from many distinct solar sources.

\subsection{$1/f$ in the Solar Wind}

\begin{figure}
\centering\includegraphics[angle=0,width=\columnwidth]{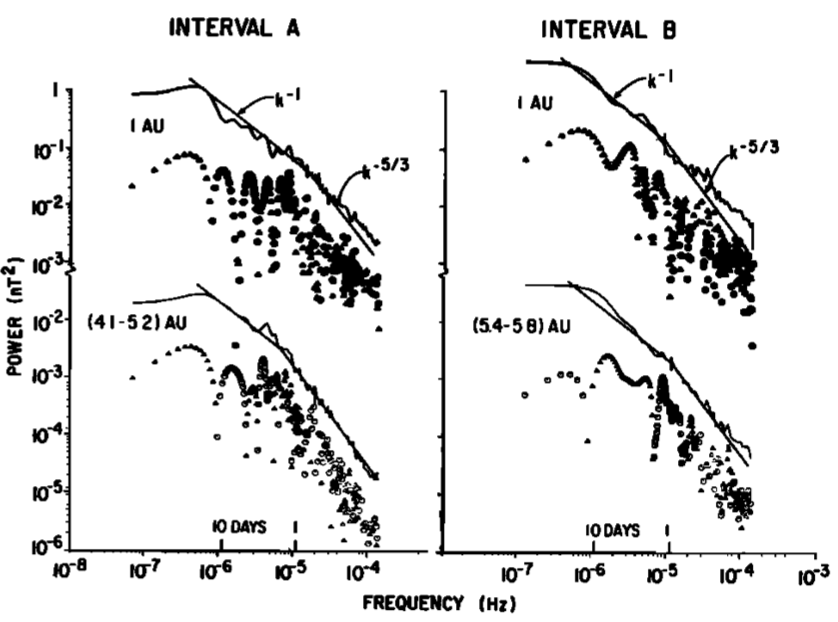}
    \caption{Magnetic field trace spectra observed at 1 au from IMP 8 and ISEE 3 and near 4 to 5 au from Voyager 1, shown as solid curves~\citep[][Fig.~2]{Burlaga84}. Triangles and circles represent positive and negative values of reduced helicity spectrum, each multiplied by frequency.}
    \label{fig:Burlaga}
\end{figure}

One of the earliest observations of $1/f$ spectrum in the solar wind dates back to \citet{Bavassano82}, where it appears in the individual components of the low-frequency magnetic field spectrum near 0.3 au, as measured by the Helios 2 spacecraft. Later, \citet{Burlaga84} find $1/f$ spectrum in the trace of the magnetic field spectral tensor across more than an order of magnitude in frequency before the correlation scales, as measured at 1 au by IMP 8 and ISEE 3, as well as near 4 to 5 au by Voyager 1 (Fig.~\ref{fig:Burlaga}). 

\begin{figure}
\centering\includegraphics[angle=0,width=\columnwidth]{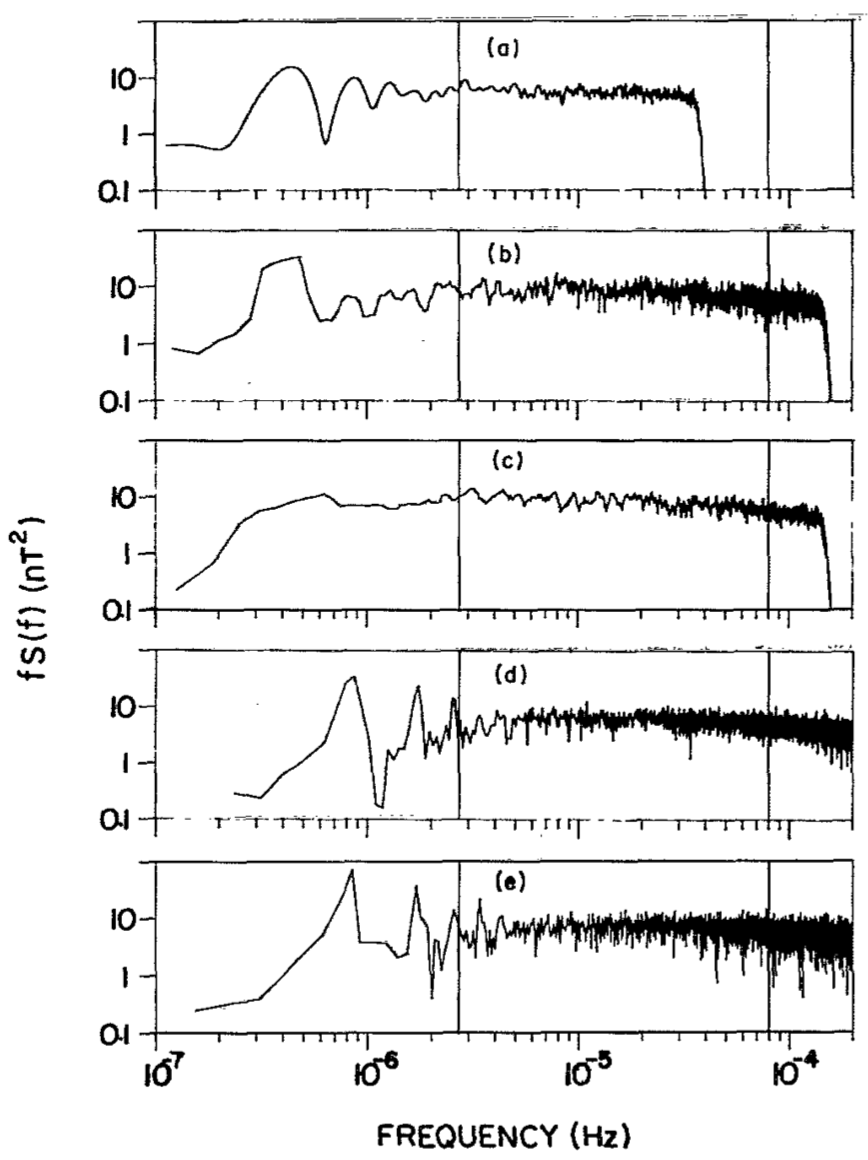}
    \caption{Compensated spectra of interplanetary magnetic field at 1 au using five intervals of data from ISEE 3 (panels a and b), IMP (panel c), and OMNI (panel d and e). $1/f$ behavior is observed between around $\unit[2\times 10^{-6}]{Hz}$ and $\unit[10^{-4}]{Hz}$~\citep[][Fig.~1]{Matthaeus86}.}
\label{fig:Matthaeus}
\end{figure}

\citet{Matthaeus86} observe $1/f$ noise within the frequency range of $2.7 \times 10^{-6}$ to $\unit[8.5 \times 10^{-5}]{Hz}$ in the 1 au magnetic field spectrum (see Fig.~\ref{fig:Matthaeus}). The authors attribute it to the superposition of signals from uncorrelated magnetic reconnection events occurring in the corona near the solar surface, with their respective correlation times collectively following a log-normal distribution. This explanation relies on the general superposition principle in which the detailed properties of the individual reconnection events are not crucial (see Section~\ref{sec:3}). We should note here that the explanation of $1/f$ provided in \citet{Matthaeus86} is easily adapted to processes other than successive coronal reconnection events, or even processes occurring beneath the photosphere.

In exploring the heliocentric dependence of the magnetic field, \citet{Bruno09} and \citet{Bruno13} observe $1/f$ scaling in the trace of the magnetic spectrum in the fast solar wind at multiple distances, using data from Helios 2 (0.3 to 1 au) and Ulysses (1.4 and 4.8 au). The upper boundary of the $1/f$ range increases in frequency as heliocentric distance decreases, from approximately $\unit[10^{-4}]{Hz}$ at 4.8 au to $\unit[10^{-3}]{Hz}$ at 0.9 au and eventually to $\unit[10^{-2}]{Hz}$ at 0.3 au. No $1/f$ signature is found in the slow wind, though it is found in a subsequent study by \citet{Bruno19} employing long-interval data.

\begin{figure}
\centering\includegraphics[angle=0,width=\columnwidth]{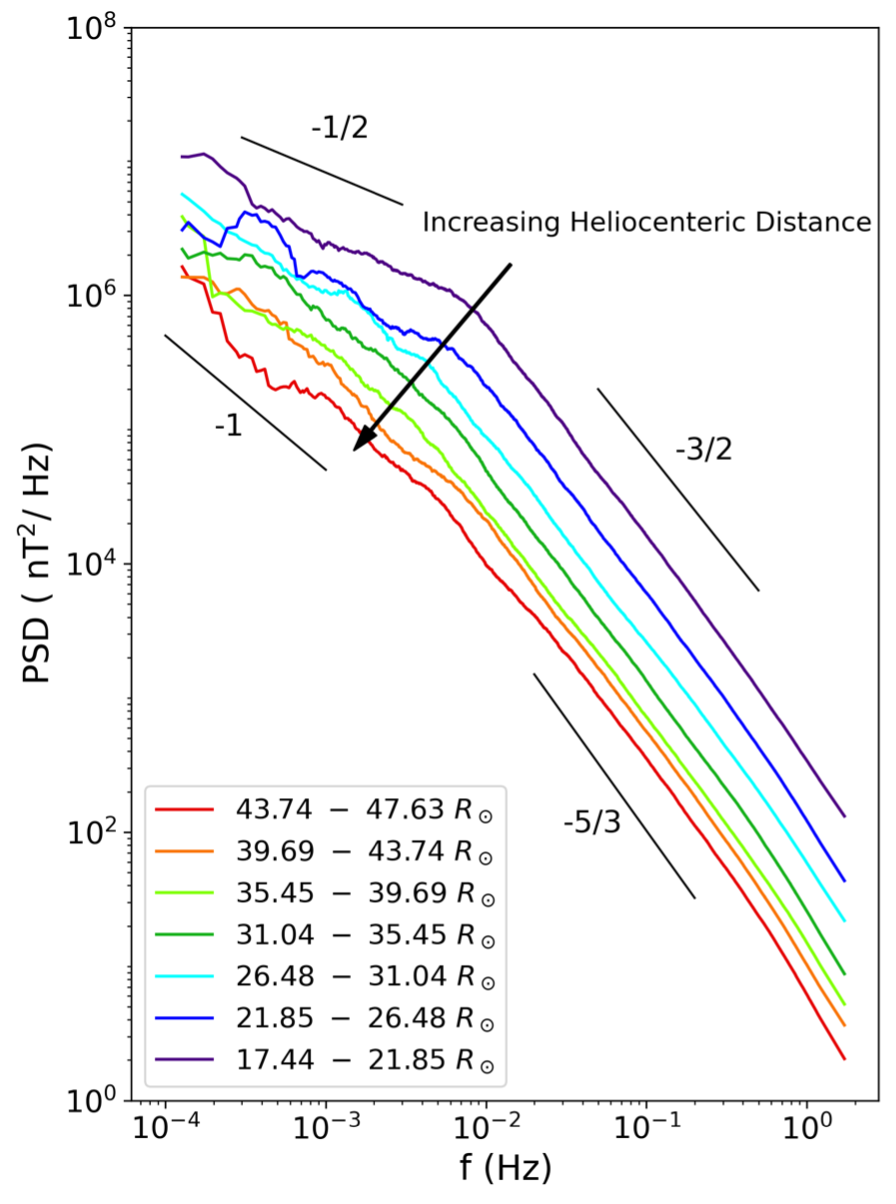}
    \caption{Magnetic field spectra of a single fast-solar-wind stream at different heliocentric distances, using data from PSP~\citep[][Fig.~3]{Davis23}.}
\label{fig:Davis}
\end{figure}

In a more recent study, \citet{Davis23} analyze magnetic field measurements in the fast solar wind streams using PSP Encounter 10 measurements and observe a spectral index of $-1$ within the energy-containing range at distances beyond around 25 solar radii. The spectral index, however, flattens to $-1/2$ closer to the Sun (see Fig.~\ref{fig:Davis}). The ``break frequency'', which serves as the approximate boundary between energy-containing and inertial scales, is also found to increase with decreasing heliocentric distance. In a companion paper, \citet{Huang23} also analyze PSP magnetic data from Encounters 1-13, focusing specifically on intervals of solar wind with nearly constant magnetic magnitude, and find spectral indices shallower than $-1$ dominating the low-frequency spectrum. However, the spectral index approaches $-1$ with increasing wind advection time and decreasing wind speed while exhibiting no dependence on heliocentric distance. These findings suggest a dynamical and possibly radial evolution of the observed $1/f$ noise in the solar wind, which favors {\it in situ} generation mechanisms within the local wind. This type of mechanism, as opposed to the non-local superposition principle, could involve a linear instability or a modification to the reasoning that leads to a Kolmogorov-like cascade~\citep{Velli89, Verdini12, Matteini18, Chandran18}. We elaborate such ideas in Section~\ref{sec:4}.

The puzzle of the origin of interplanetary $1/f$ noise presents a clear but fundamental dichotomy that is not yet unambiguously resolved -- does this signal emerge from a collective statistical principle~\citep[as suggested by][]{Matthaeus86}, or is it a consequence of a particular dynamical process emerging from local dynamics alone~\citep[as suggested by][]{Davis23, Huang23}? In the following section, we review theories relevant to the former argument. Then in Section~\ref{sec:4}, we provide an in-depth examination of solar wind observations and related simulation and experimental results, which may shed light on this ongoing debate.

\section{A Robust Superposition Principle}
\label{sec:3}

Theories have been established concerning a generic generation of $1/f$ noise through a superposition principle. \citet{Machlup81} proposes that if samples of ``purely'' random processes (those with autocorrelations of the form $e^{-t/\tau}$ where $\tau$ is the correlation time of the sample) exist with a minimum correlation time $\tau_1$ and a maximum $\tau_2$, then a scale-invariant distribution of the correlation times would lead to an averaged power spectrum exhibiting $1/f$ scaling over the frequency range $1/\tau_2 \ll \omega \ll 1/\tau_1$. The resulting $1/f$ spectrum can span several orders of magnitude if $\tau_2$ is much greater than $\tau_1$. However, in the context of the solar wind, two notable deviations from Machlup's model apply: (1) correlation times tend to follow log-normal distributions instead of inverse distributions~\citep[see, e.g.,][]{Ruiz14, Isaacs15}, and (2) measurements typically display a Kolmogorov power-law spectrum with an index of \(-5/3\) in the inertial range (on scales smaller than the correlation scales) under the regime of incompressible, isotropic magnetohydrodynamic (MHD) turbulence with high Reynolds numbers~\citep[see, e.g.,][]{Matthaeus07, Bruno19}. \citet{Montroll82} show that a log-normal distribution with a sufficiently large variance has an extended, inverse-like proportion. We also show in Appendix~\ref{app:1/f} that an ensemble of datasets with an arbitrary power-law index $-\alpha < -1$ can collectively produce a $1/f$ spectrum through the same sense of superposition.

\subsection{Machlup's Superposition Principle}
\label{subsec:Machlup}

In examining the relationship between $1/f$ noise and its mechanism of generation, necessarily involving occurrences of unexpected or rare events,\footnote{Events with scale-invariant occurrences are often considered unexpected because the tail of the distribution is asymptotic and non-integrable. So tail events are inherently rare but always possible, making them unexpected.} \citet{Machlup81} hypothesizes that if the spectrum has no characteristic time -- wherein the same amount of energy resides between any two frequencies separated by a fixed factor -- then so might the distribution of the generating events. Suppose there exists an ensemble of nonlinear, chaotic processes naturally displaying exponentially decaying autocorrelations, and their characteristic correlation times $\tau$ follow the scale-invariant (or inverse) distribution 
\begin{equation}
    \rho(\tau) d \tau \propto \frac{d \tau}{\tau}.
\label{eq:inverse}
\end{equation}
Then the spectrum of an individual event, as written below in angular frequency ($\omega = 2\pi f$) domain, is of the form of a Lorentzian:
\begin{equation}
    S(\tau, \omega) \propto \int_{-\infty}^\infty e^{-i\omega t} e^{-t/\tau} dt \propto \frac{\tau}{1+\omega^2 \tau^2}.
\label{eq:Stau}
\end{equation}
And the overall spectrum of this ensemble of events of equal integrated power\footnote{Derivation of \citet{Machlup81} assumes that all samples in the ensemble share the same variance, or integrated power across scales. However, this condition is not necessary -- $1/f$ spectrum can still emerge following the logic of Eq.~\ref{eq:Sbar} if the variances follow a scale-invariant distribution.} is
\begin{equation}
    \overline{S}(\omega) = \int_{\tau_1}^{\tau_2} S(\tau, \omega) \rho(\tau) d{\tau} \propto \frac{\tan^{-1}(\tau \omega)}{\omega} \Big|_{\tau_1}^{\tau_2},
\label{eq:Sbar}
\end{equation}
where $\tau_1$ and $\tau_2$ are the minimum and maximum correlation times of the generating events, respectively. If the correlation times span several orders of magnitudes, i.e., $\tau_1 \ll \tau_2$, then within the frequency region satisfying $\tau_1 \ll 1/\omega \ll \tau_2$, 
\begin{equation}
    \overline{S} (\omega) \propto \frac{\pi/2 + \mathcal{O}(\tau_2 \omega)^{-1} + \mathcal{O}(\tau_1 \omega)^1}{\omega}.
\end{equation}
To a zeroth-order approximation, $\overline{S} (\omega) \propto 1/\omega$.

\subsection{Connection Between Inverse and Log-Normal Distributions}

We note that the correlation time distribution as given in Eq.~\ref{eq:inverse} is not integrable, neither are the sharp boundaries of $\tau_1$ and $\tau_2$ assumed in the last section physical. So what exactly happens in the limits as $\tau \rightarrow 0$ and $\tau \rightarrow \infty$? \citet{Montroll82} propose the log-normal distribution that behaves like $1/\tau$ in the intermediate range and naturally governs multiplicative processes in which the total probability is the product of probabilities of several independent random variables.

A log-normal distribution describes a random variable $X$ whose logarithm is normally distributed. Suppose $Z$ is a standard normal variable. Then a log-normal distribution is defined for variable $X = e^{\mu + \sigma Z}$ as 
\begin{equation}
    f(x) = \frac{1}{x \sigma \sqrt{2\pi}} \exp{\left[ - \frac{\left(\log{x}-\mu \right)^2}{2\sigma^2} \right]},
\label{eq:lognormal}
\end{equation}
where $\mu$ and $\sigma$ is the mean and standard deviation of the variable $\log{X}$, respectively. From Eq.~\ref{eq:lognormal}, an inverse scaling $f(x) \sim 1/x$ under certain conditions becomes apparent. If we define $\overline{x} \equiv e^\mu$, then $f(x) \propto 1/x$ when 
\begin{equation}
    \frac{\left[ \log(x/\overline{x}) \right]^2}{2 \sigma^2} \ll 1.
\end{equation}
Following conventions from \citet{Montroll82}, given a fraction of tolerable deviation $\theta$, a log-normal distribution is scale-invariant within a fraction of $\theta$ inside the domain of random variable $x$ satisfying\footnote{The condition in \citet{Montroll82} is $|\log{(x/\overline{x})}| \leq 2 \theta \sigma^2$, which is established in log scale. The condition as in Eq.~\ref{eq:inverse_constraint} is established in linear scale.}
\begin{equation}
    \left| \log{\left( \frac{x}{\overline{x}} \right)} \right| \leq \sqrt{2\theta \sigma^2}.
\label{eq:inverse_constraint}
\end{equation}

It is known that many solar wind variables, such as the correlation times, solar wind speed, density, temperature, magnetic field strength, and so on, follow log-normal distributions~\citep[see, e.g.,][]{Burlaga99, Padhye01, Ruiz14, Isaacs15}.\footnote{We note that a detailed analysis by \citet{Feynman94} finds potential departure from log-normality in hourly-averaged magnetic fluctuations at 1 au. However, the distribution of 3-hour averaged magnetic magnitudes is found to be close to log-normal. A recent study by \citet{Huang24} employing PSP data reports that after removing power-law radial dependence, intervals of magnetic magnitudes displaying Gaussianity permeate the Alfv\'enic solar wind.} The abundance of log-normal distributions may be attributed to the multiplicative nature of physical processes occurring within the solar wind. Suppose the value of a random variable $X$ is based on the product of $N$ independent variables $X_1, \cdots, X_N$, i.e.,
\begin{equation}
    X = X_1 \times \cdots \times X_N.
\label{eq:X}
\end{equation}
Then evoking the Lyapunov central limit theorem, the distribution of the logarithm, $\log{X} = \log{X_1} + \cdots + \log{X}_N$, follows a normal distribution if certain regular conditions are satisfied. In \citet{Shockley57}, $X$ may represent the productivity of a researcher, while $X_{1, \cdots, N}$ represent research merits, such as the ability to recognize a research topic, the ability to evaluate results, and so on. 

In the case of the solar wind, $X_{1, \cdots, N}$ may represent successive reconnection events or successive foldings in a dynamo~\citep[see, e.g.,][]{Matthaeus86, Bourgoin02, Ponty04, Dmitruk07}. In particular, magnetic structures of initial lengthscale $\lambda_0$ may experience $N$ successive reconnections, each modifying $\lambda_0$ by a factor of $(1+\epsilon)$, so that the ensemble lengthscale $\lambda = \lambda_0 (1+\epsilon)^N$ mimics Eq.~\ref{eq:X} and is log-normally distributed~\citep{Matthaeus86}. In such cases where the successive events $X_{1, \cdots, N}$ are identical, the Lindeberg–L\'evy central limit theorem dictates that $\log{X}$ is normally distributed for sufficiently large $N$ if the second moment of $\log{X_{1, \cdots, N}}$ exists.

\section{Theoretical Issues and Further Observations}
\label{sec:4}

The superposition principle elaborated upon in Section~\ref{sec:3} provides a compact pathway to explain the generation of $1/f$ signals, but it is not specific regarding the origin of the underlying processes that are superposed. Therefore any hypothesis concerning the nature of these processes must be assessed for consistency based on principles beyond the superposition mechanism itself. On this basis, the suggestion by \citet{Matthaeus86} that the superposition involves successive reconnections occurring in the deep corona avoids problems related to the available time scales. Specifically, sub-Alfv\'enic coronal dynamics (including reconnections) are not strongly limited by convection time or expansion time, since in this region the MHD characteristics travel both toward and away from the Sun.

While there has been a variety of mechanisms proposed to explain shallow $1/f$ solar wind spectra, especially in the trace magnetic field component spectra, these mechanisms are usefully categorized into those that operate {\it locally} in the interplanetary medium and those that originate in the lower corona or even in the solar interior. From a practical and physical point of view, the region for local processes can be viewed as the super-Alfv\'enic solar wind, whereas coronal and solar processes operate at lower altitudes. We distinguish local and solar mechanisms for generating the $1/f$ signal in this simple and perhaps {\it ad hoc} way.  

\subsection{Local Origins of $1/f$.}

The pioneering observations by \citet{Burlaga84} set the stage for wide ranging investigations into the origins of the interplanetary $1/f$ signal. It is noteworthy that their Fig.~2 (Fig.~\ref{fig:Burlaga} here) is labeled by both frequency and wavenumber, where the wavenumber spectrum assumes the form $k^{-1}$, in accordance with the Taylor frozen-in hypothesis~\citep{Taylor38}. 

Associated with this spectral law, \citet{Burlaga84} offered as a possible explanation for the observed $1/f$ an inverse cascade of magnetic helicity. We evaluate this as a suggestion of a local process. (We revisit it later as a possible solar process.) The inverse cascade process operates in the regime of incompressible magnetohydrodynamics~\citep[MHD; see][]{Frisch75, Montgomery78} under conditions where helicity is an ideal invariant. In freely decaying turbulence with appropriate boundary conditions, magnetic helicity is known to be important and can be responsible for effects such as Taylor relaxation~\citep{Taylor74} or selective decay~\citep{Montgomery79}. Although these mechanisms are intriguing and physically appealing, their applicability to the solar wind is questionable, as the magnetic helicity of 
fluctuations is not an ideal invariant in magnetohydrodynamics (MHD) when a mean field is present -- such as the Parker spiral field that threads the interplanetary medium. In addition, inverse cascade processes are generally very slow compared to direct cascade processes, and the observed $1/f$ range at 1 au has only a few nonlinear times to develop during transit from the Sun~\citep{Zhou90}. In any case the inverse cascade scenario involves {\it local} MHD scale processes that would need to occur in transit from the corona if the process is to be considered local. 

On the other hand, the case where $1/f$ signal is already present below the Alfv\'en critical region is unrelated to {\it in situ} solar wind dynamics and avoids transit time issues. \citet{Burlaga84} mention the possibility of an alternative explanation that would involve ``some appropriate superposition of streams''. This suggestion was subsequently developed into the \citet{Matthaeus86} model, which employed the superposition principle described in Section~\ref{sec:3}.

Another theoretical approach to explaining the solar wind's $1/f$ spectrum as a local process was proposed by \citet{Velli89} and \citet{Verdini12}. Their model differs from the superposition principle in that the physical processes producing the $1/f$ signal occur {\it in situ} in the evolving solar wind due to the interaction of Kolmogorov-like nonlinear effects and the influence of expansion. Ideas along these lines have been examined in recent Parker Solar Probe (PSP) observations~\citep{Davis23, Huang23}, though remote generation in the corona is not ruled out by these authors. A variant of the above ideas has appeared in a recent preprint~\citep{Huang24arXiv}.

These recent studies based on PSP observations discuss their results in the context of the models proposed by \citet{Chandran18} and \citet{Matteini18}. \citet{Chandran18} uses weak-turbulence theory to suggest that the \(1/f\) behavior can arise from the parametric decay of Alfv\'en waves that are initially highly imbalanced (with a dominant outward propagating mode), leading to an inverse cascade wherein the dominant Alfv\'en mode acquires a \(1/f\) scaling over time. A narrowing of the range of frequencies displaying \(1/f\) is then predicted for heliocentric distances smaller than 0.3 au. A similar prediction was made by \citet{Matteini18} based on arguments relating to the low magnetic compressibility of the solar wind. The authors conjecture that \(1/f\) is the steepest possible spectrum at (large) scales where \(\delta B/B\sim 1\), a limit imposed by the saturation of magnetic fluctuations \(\delta B\) that are bounded on a sphere of radius equal to the background field \(B\). Consequently, near the Sun, where \(\delta B/B < 1\)~\citep[see, e.g.,][]{Chhiber22}, the \(1/f\) range is predicted to disappear. See \citet{Huang23} for a detailed discussion on how models by \citet{Chandran18} and \citet{Matteini18} are used to interpret recent PSP observations.

\subsection{Causality and Range of Influence}
\label{sec:causality}

A feature of all the local mechanisms presented so far is that they refer only to observed frequencies above around $\unit[10^{-4}]{Hz}$~\citep[see, e.g.,][]{Huang23, Davis23}. However, as emphasized above, the entirety of the observed interplanetary $1/f$ signal extends to frequencies as low as $\unit[2 \times 10^{-6}]{Hz}$.\footnote{In fact, it is possible that the signal extends to even lower frequencies. However, the signal tends to merge with harmonics of the solar rotation period, as seen in Fig.~\ref{fig:Matthaeus}.} \citet{Huang23} and \citet{Davis23} study the tendency toward slopes shallower than the inertial range values, and even shallower than $1/f$, when examining frequencies between $10^{-2}$ and $\unit[10^{-4}]{Hz}$, a phenomena noted in similar frequency ranges in early works~\citep{Bavassano82, Denskat82} but without clear explanation. Without prejudice to the applicability of the local mechanisms down to $\unit[10^{-4}]{Hz}$, we may ask if such processes can be extended to much lower frequencies. Nearly two more decades of frequency must be accessed to include the full observed $1/f$ range.

The concept of causality or ``range of influence'' becomes critical at this juncture~\citep{Zhou90, Chhiber2018thesis, Matthaeus2018AGU}. We ask the question: Over what distance can MHD processes exert influence during passage to 1 au? An estimation of this distance amounts to a position-dependent estimate of an MHD causality limit using a few key time and space scales. Start by assessing the maximum distance an MHD signal can travel in transit to $\unit[1]{au}= \unit[1.5 \times 10^{13}]{cm}$. The transit time for the wind at $\unit[400]{km/s}$ is then $T_\mathrm{tr} \sim \unit[3.8 \times 10^5]{s}$. An upper bound estimate of range of influence using the Alfv\'en speed ($\sim \unit[50]{km/s}$) is $L_\mathrm{roi} = V_A T_\mathrm{tr} \sim \unit[0.1]{au}$. This corresponds to a time of passage $T_\mathrm{roi} = L_\mathrm{roi} /V_\mathrm{SW} \approx \unit[3.5 \times 10^4]{s}$ and a frequency $f_\mathrm{roi} = 1/T_\mathrm{roi} = 2.8 \times \unit[10^{-5}]{Hz}$. This is the lowest frequency that can be influenced by Alfv\'en wave propagation at 1 au. For turbulent motions propagating at speed $\delta V < V_A$, the range of influence will decrease, and the corresponding frequency will be higher than $f_\mathrm{roi}$. The fiducial line drawn in Fig.~\ref{fig:Matthaeus} is around $\unit[8 \times 10^{-5}]{Hz}$ and in the present estimation would correspond to a turbulence amplitude a factor of 2 or 3 smaller than $V_A$.

The range of influence can be meaningfully compared with other time and length scales. First, the correlation scale $L_c$ at 1 au is on average around $\unit[10^6]{km}$, with broad variation~\citep{Ruiz14}. At $V_\mathrm{SW} = \unit[400]{km/s}$, a structure of size $L_c$ passes an observation point in $\unit[2.5 \times 10^{3}]{s}$, corresponding to $\unit[4 \times 10^{-4}]{Hz}$, again with substantial variation. This is comfortably greater than the frequency associated with range of influence, as it must be. 

Next, we note that solar wind plasma streaming transits 1 au at $\unit[400]{km/s}$ in time $T_\mathrm{tr} \sim \unit[100]{hr}$, or in frequency, $f_\mathrm{tr} = 1/T_\mathrm{tr} \sim \unit[2.8 \times 10^{-6}]{Hz}$. Comparing these with observations (e.g., Fig.~\ref{fig:Matthaeus}), we see that 
the $1/f$ noise range rather neatly spans (and extends somewhat beyond) the frequency interval from $f_\mathrm{tr}$ to $f_\mathrm{roi}$. 

The fact that the nominal correlation frequency is somewhat higher than the range of influence frequency is reasonable and expected. The spectrum is well known to roll over into $\sim f^{-5/3}$ at frequencies above $f_\mathrm{cor}$. However, since the correlation lengths in individual samples are known to be log-normally distributed~\citep{Ruiz14}, this rollover is not expected to be sharp. Moreover, as stated above, the range of influence differs in turbulence with different amplitudes $\delta V$, and this amplitude itself is broadly distributed, perhaps again in a log-normal fashion~\citep{Padhye01}. Finally the approximation that the log-normal distribution of correlation times is nearly scale-invariant may break down at the extremes of the $1/f$ spectral range. Given these variabilities, it seems inevitable that the transition from the $1/f$ range to the turbulence $f^{-5/3}$ range takes place gradually. In observations such as Figs.~\ref{fig:Burlaga},~\ref{fig:Matthaeus}, and~\ref{fig:Davis}, this transition takes place over about a decade of frequency. 

Consideration of the above timescale inequalities implies that a Kolmogorov-like direct cascade, or any process requiring standard nonlinear turbulence time scales, will not be able to operate over the full range of the observed $1/f$ spectrum in the time of available transit to 1 au. Analogous estimates can be readily made for other heliocentric distances or other values of turbulence timescales. This casts doubt on theoretical explanations for the $1/f$ signals. Minimally it indicates that {\it in situ} explanations must be supplemented by some other non-local process that fortuitously produces a spectrum that smoothly matches the spectrum that extends to much lower frequencies. 

\subsection{Origin in Coronal and Solar Processes}

\begin{figure}
\centering\includegraphics[angle=0,width=\columnwidth]
{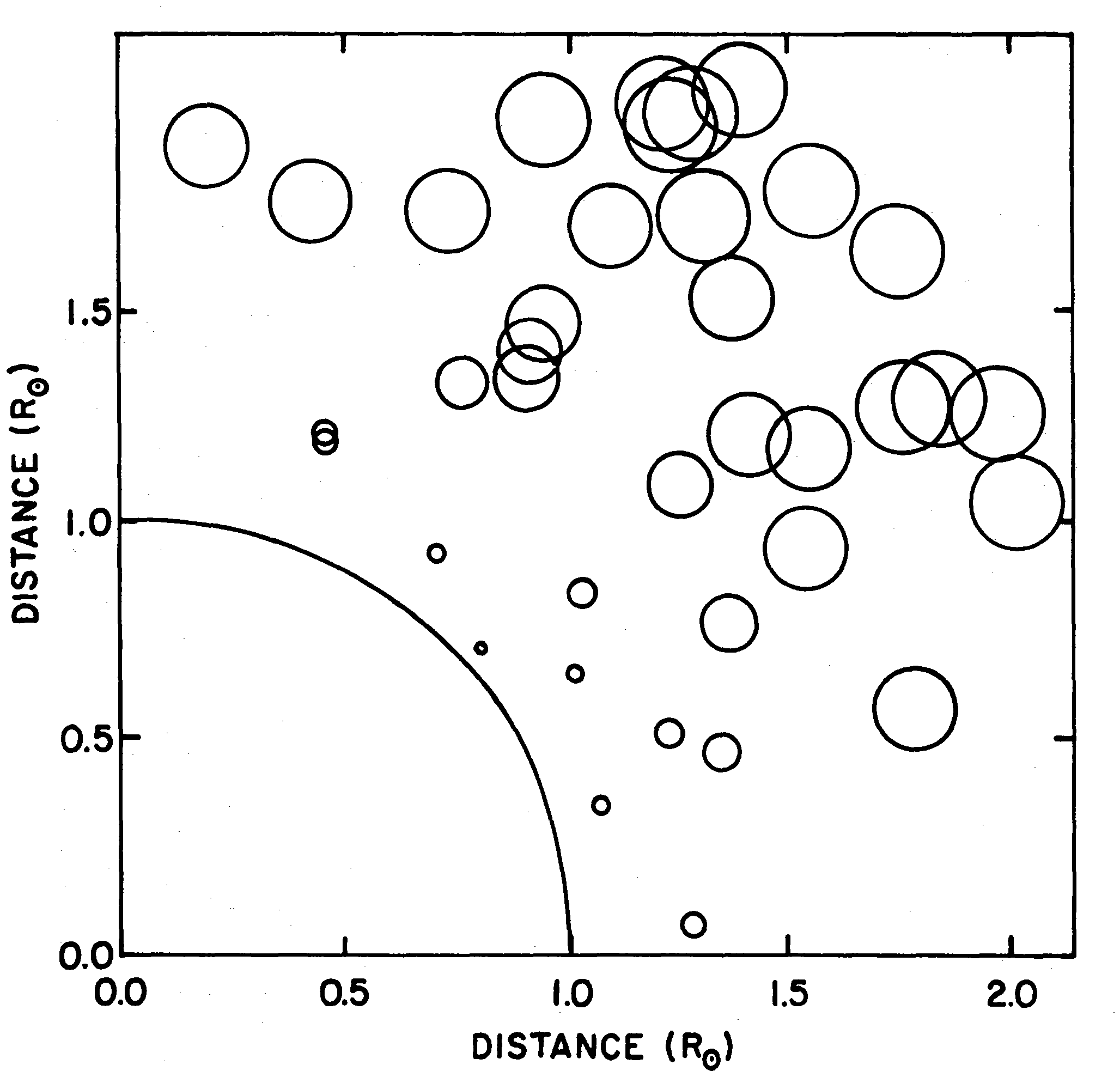}
    \caption{Cartoon of successive scale-invariant reconnection model as interpreted by \citet{Mullan90}.
  }
\label{fig:Mullan}
\end{figure}

In conjunction with $1/f$ observations from long data records at 1 au, \citet{Matthaeus86} offered a theoretical explanation based on a particular scenario in which the Montroll-Shlesinger-Machlup superposition principle (see Section~\ref{sec:3}) is invoked. It is based on the elementary idea that a collision between two flux tubes can lead to reconnection and merging, producing a plasmoid with a larger cross section, potentially doubling in size if the colliding flux tubes are of equal dimensions. This process is repeatable given a certain probability for reconnection and merger. Several stages of such merger lead to a multiplicative process and hierarchy that may be described using a log-normal distribution. Then, invoking the developments of \citet{Montroll82}, a scale-invariant distribution can be achieved over some range of scale sizes. As these structures are accelerated into the solar wind, they plausibly represent the $1/f$ signal observed without further dynamical evolution. The model of \citet{Mullan90} based on photospheric and coronal observations added considerable detail to the successive, scale-invariant coronal reconnection model. A cartoon taken from \citet{Mullan90} is shown here as Fig.~\ref{fig:Mullan}. 

The concept that coronal reconnection is a multiscale process capable of supporting numerous successive reconnections is well founded in observations, based on the detection of the low-lying mixed polarity magnetic carpet~\citep{Schrijver03}. Simulations have shown that prolific flux tube collisions and resulting current sheet formation are to be expected in this highly dynamic, anisotropic turbulent medium~\citep[see, e.g.,][]{Einaudi94, Servidio09}. 

The successive reconnection scenario readily lends itself to a description of a multiplicative process in the sense of Eq.~\ref{eq:X}. The expectation of a log-normal distribution of the resulting correlation scales follows immediately, and in fact, such a distribution is observed~\citep[see, e.g.,][]{Ruiz14, Isaacs15}.

Another mechanism operating in the corona has been suggested by \citet{Magyar22}, based on the coupling of Alfv\'en waves with a transversely inhomogeneous background density field. Through 3D compressible MHD simulations, in which transverse velocity perturbations are injected at the lower boundary, the authors demonstrate that the resulting magnetic field line oscillations exhibit coherent behavior, leading to the formation of transverse collective modes, such as surface Alfv\'en waves. This occurs when the coupling between the waves and the inhomogeneous background is fully considered. Under linear processes like phase mixing and resonant absorption, these surface Alfv\'en waves develop a perpendicular energy spectrum that mirrors that of the background density. As a result, an inhomogeneous density field with a perpendicular $1/f$ spectrum can give rise to a $1/f$ spectrum in the perpendicular Alfv\'en velocity fluctuations. However, the authors acknowledge that the spectral index of density field in the solar corona remains uncertain, with values reported to range from $-1$ to $-2$~\citep{Woo79, Moncuquet20, Carley21}. Additionally, it is unclear how this process relates to the photospheric magnetic structure and features of the solar dynamo, as we discuss in Section~\ref{sec:dynamo}.

\subsection{Spectral Evolution and Generation of Correlation} 

The above scenario is essentially a coronal process that does not require the participation of {\it in situ} dynamics in the super-Alfv\'enic wind. However, as the solar wind expands, local turbulence can generate spatial correlations in the form of the Kolmogorov spectrum and its hierarchy of {\it locally produced} magnetic field and vorticity structures~\citep{Bruno13}. A familiar and often quoted {\it net} product of such nonlinear couplings is a net flux of energy through the inertial range toward smaller scales. However, this {\it cascade} is the average result of a vast number of triadic interactions~\citep{Verma19}, which transfer energy almost equally toward larger and smaller scales, with the net being toward the latter.

This property has numerous influences on a turbulent system, but relevant here is the expectation that freely evolving turbulence will generate correlations at increasing scales. This is a fundamental feature of the \citet{deKarman38} analysis of homogeneous turbulence, and it implies that the similarity scale that defines the long-wavelength bound on the inertial range must increase in time. In the present context, this leads to the observed increase in length scale of the solar wind correlations on average with increasing radial distance~\citep{Klein92, Ruiz14}. This gradually converts the observed high frequency part of the $1/f$ range into the correlated Kolmogorov-like $f^{-5/3}$ (or $k^{-5/3}$ using the frozen-in property) range. Indeed, this is observed and well documented~\citep{Tu95, Bruno13} as the upper end of the $1/f$ range evolves toward lower frequency at increasing radial distance. This has been often described as the migration toward lower frequency of the ``break point'' between the Kolmogorov and $1/f$ spectral ranges~\citep{Bavassano82, Tu84, Tu95, Wu20, Wu21}, even if this transition is often gradual rather than sharp, as illustrated, e.g., in Fig.~\ref{fig:Davis}.

\section{Connections to Inverse Cascade and Dynamo}
\label{sec:dynamo}

The possibility that the observed $1/f$ signal is related to inverse cascade~\citep{Frisch75, Fyfe77} activity crucially depends on where the process is purported to occur. For {\it in situ} inverse cascade-related activity in the super-Alfv\'enic solar wind, the issue of available time immediately enters. The problem with establishing the observed $1/f$ spectrum in the interplanetary medium due to direct cascade processes was discussed in Section~\ref{sec:causality}. However, it is well known that inverse cascade processes are significantly slower than their direct cascade counterparts and have been been shown in various circumstances to require many, even hundreds, of nonlinear eddy turnover times to have significant effects~\citep{Dmitruk07, Dmitruk11} on low frequency time variations. On this basis, it seems that accounting for the full frequency range of the observed $1/f$ solar wind noise through inverse cascade in the super-Alfv\'enic solar wind is essentially ruled out. 

In the sub-Alfv\'enic corona and in the solar dynamo, the situation regarding inverse cascade activity is markedly different. In the low plasma-beta corona, plasma turbulence is likely characterized by a high degree of quasi-two dimensional anisotropy, often described by {\it Reduced Magnetohydrodynamics}~\citep[RMHD;][]{Kadomtsev74, Rappazzo08, Gomez13, Perez13}. Such systems asymptotically approach two dimensionality, a limit in which a {\it quasi-conserved} mean square magnetic potential can apparently support inverse cascade activity~\citep{Fyfe77, Dmitruk07}. This process is indeed associated with successive reconnections of a sea of magnetic flux tubes~\citep{Servidio10}. It is also associated with the more rapid decay of energy relative to mean square potential~\citep{Matthaeus80}, a feature known as {\it selective decay}. Therefore, it is possible to view the model for $1/f$ driven by scale-invariant coronal reconnections~\citep{Matthaeus86, Mullan90} as supported by inverse cascade processes. 

Likewise, for the solar dynamo, connections between inverse cascade and $1/f$ noise generation are equally clear. Observations provide a ``smoking gun'' involvement of sub-photospheric dynamics in producing observable spectral signatures. Particularly suggestive are observed azimuthal wavenumber spectra from photospheric line-of-sight magnetic fields reported by \citet{Nakagawa74} using Kitt Peak magnetograms and by \citet{Matthaeus07} employing data from the SOHO/MDI instrument. In each case, there is evidence of a $1/k$ dependence in the photospheric {\it spatial} structure, which may be regarded as a signature of MHD inverse cascade~\citep{Frisch75}. Elementary models provide a simple relationship between spatial structures in the photosphere and observed spectra in the solar wind~\citep{Giacalone06}.

\begin{figure}
\centering\includegraphics[angle=0,width=0.49\columnwidth]{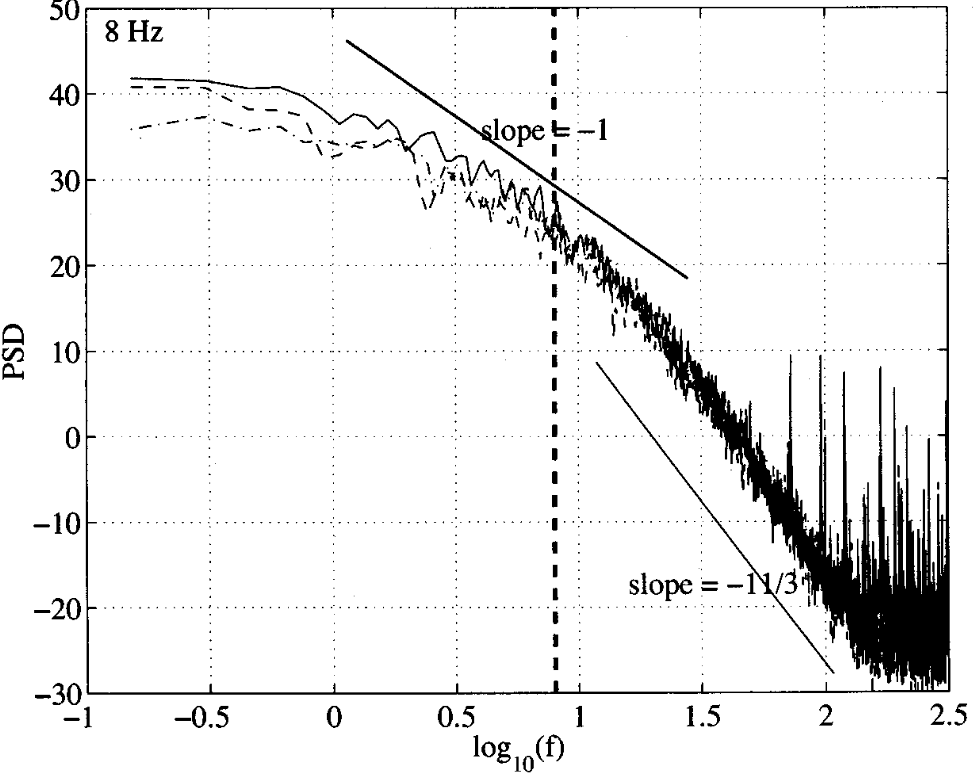}
\centering\includegraphics[angle=0,width=0.49\columnwidth]{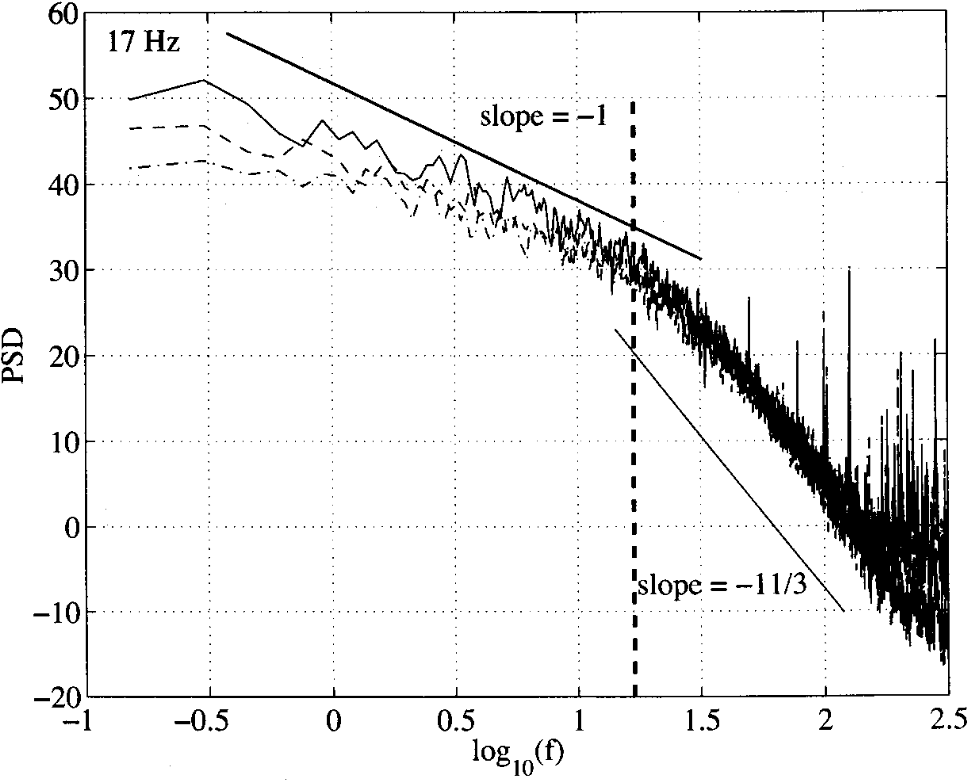}
\centering\includegraphics[angle=0,width=\columnwidth]{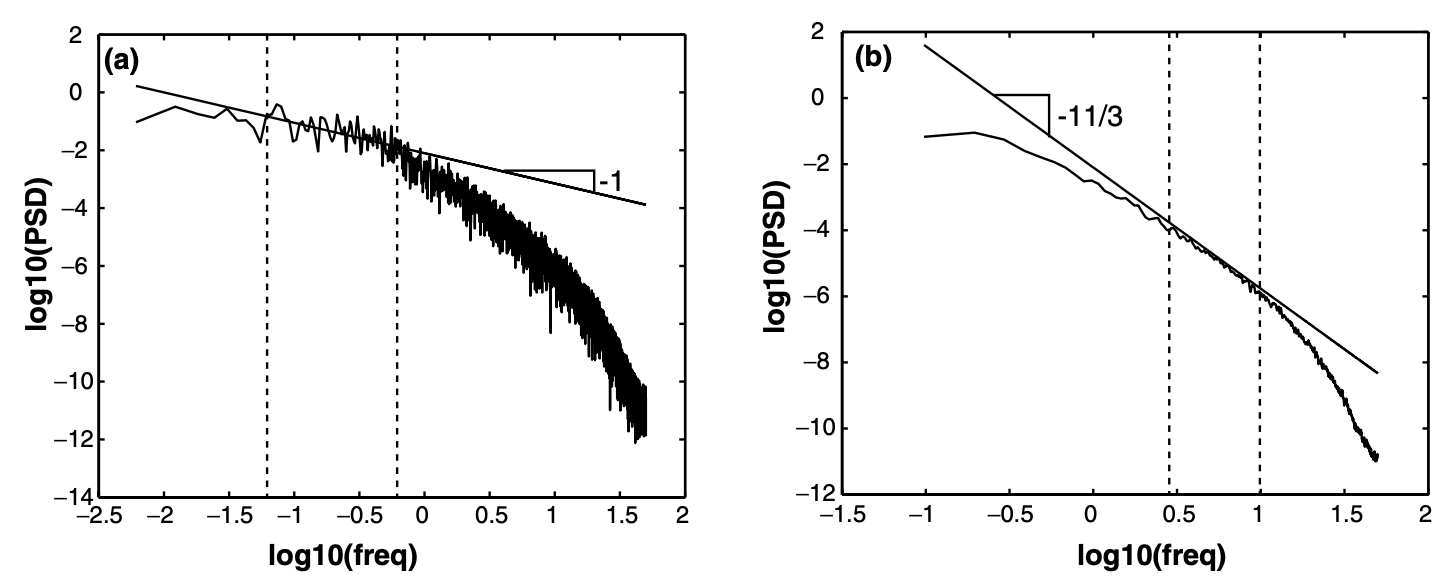}
    \caption{Top: Frequency spectra of magnetic field from liquid sodium dynamo experiment with low (left panel) and high (right panel) disk rotation rate~\citep[][Fig.~11]{Bourgoin02}. Bottom: Magnetic spectra from low magnetic Prandtl number and large kinetic Reynolds number simulation estimated over long (left panel) and short (right panel) intervals. $1/f$ appears in frequency domain between $1/t_0$ to $1/(10t_0)$ where $t_0$ is the magnetic diffusion timescale~\citep[][Fig.~4]{Ponty04}.}
\label{fig:Ponty04}
\end{figure}

On the theoretical side, spherical MHD dynamo simulations carried out for very long time scales indicate spectral transfer consistent with inverse cascade while also producing $1/f$ noise in the time domain. Such simulations employ highly idealized boundary conditions and ideal MHD equations~\citep{Dmitruk14}, thus enabling long timescale runs. The $1/f$ signals appear when runs are initialized with significant magnetic helicity or with significant rotation. In the same instances, the flows experience strong condensation of energy into the largest scale degrees of freedom in the sphere, a signature of the possibility of inverse cascade~\citep[as in][]{Frisch75}. There is also support for dynamo generation of $1/f$ in laboratory experiments~\citep{Bourgoin02, Gailitis04} and supporting simulations~\citep[see][and Fig.~\ref{fig:Ponty04}]{Ponty04, Ponty05}.

\section{Observability by PUNCH}

Remarkably, and perhaps fortuitously, the PUNCH mission~\citep{DeForest22} will provide images covering a range of scales arguably of direct relevance to the observed $1/f$ spectra. PUNCH's high-resolution imaging data is designed to have at least \(10\times\) better resolution than previous imagers such as STEREO~\citep{DeForest16, DeForest18}, thus in principle resolving structures at scales within the turbulence inertial range. Roughly speaking, PUNCH images will span scales up to a few 1 au, and with resolutions as fine as $\unit[10^{6}]{km}$, comparable to the expected correlation scale. Therefore PUNCH will in principle capture images of the plasma responsible for the $1/f$ signal. The challenge lies in the interpretation of the images. 

\begin{figure}
\centering\includegraphics[angle=0,width=\columnwidth]{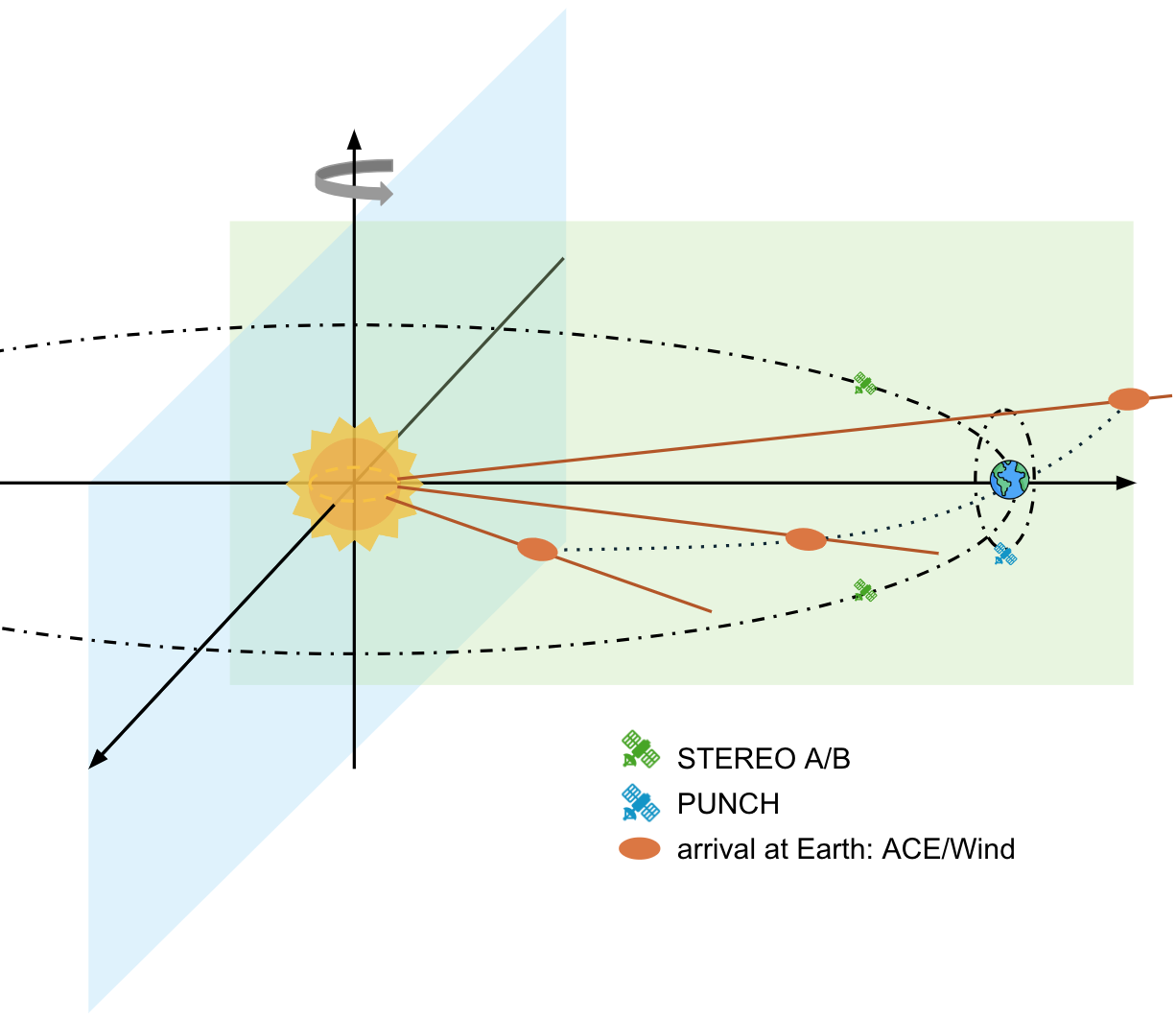}
    \caption{Projected fields of view (FOV) of STEREO A/B (green plane) and PUNCH (blue plane), both perpendicular to the ecliptic plane. Wind parcels that have reached or will reach Earth as observed by ACE/Wind are labeled in orange, with the simple assumption that the solar wind corotates rigidly with the Sun.}
\label{fig:schematic}
\end{figure}

One issue is that PUNCH will detect density variations, not magnetic field, except perhaps by inference with regards to coronal structures~\citep[see, e.g.,][]{DeForest16,DeForest18,Ruffolo20}. However, there are some {\it in situ} observations of $1/f$ signal in density~\citep{Matthaeus07}, as well as inferred results from the SOHO UVCS instrument in the deep corona~\citep{Bemporad08}. PUNCH will also observe the inner solar wind and corona in a different orientation than either STEREO imaging, or {\it in situ} measurements such as those from the ACE or Wind missions at a fixed position using the Taylor hypothesis. Fig.~\ref{fig:schematic} illustrates the essence of these differences in a highly idealized format. The most significant issue in quantifying the correspondence between PUNCH images and spectral characteristics is the averaging over depth of field that is intrinsic in PUNCH images. This greatly complicates the interpretation of image spatial scales to the {\it in situ} observed range of $1/f$ frequencies. 

Preliminary studies of this problem have begun in anticipation of the PUNCH launch~\citep[see][]{Pecora24}. Efforts so far have adopted a constructive procedure in which data from magnetohydrodynamic numerical simulations of turbulence are subject to modeling to in effect synthesize PUNCH data. This takes into account not only the line-of-sight integration, but also the three dimensional response due to Thomson scattering of white light from the heliospheric density variations. Progress in these studies have shown that standard turbulence spectral scalings are modified in the synthetic PUNCH images, even if there are promising indications that information about the original properties of the turbulent field remains present. Such studies are first steps in understanding whether PUNCH will permit quantitative diagnosis of the presence of the low-frequency $1/f$ signals. Provided that ongoing studies can establish useful connections between PUNCH images and spectral distributions, this mission will have the potential to reveal the origin and evolution of the elusive interplanetary $1/f$ signal.

\section{Discussion}

$1/f$ noise appears in such a wide variety of physical systems~\citep{Dutta81} that it is difficult to fully review its occurrences in a limited space. Nevertheless, we attempt here to provide a broad but incomplete survey to suggest 
its generic nature. There are also a variety of detailed mechanisms that are suggested to explain its emergence. We have not delved deeply into these in systems other than the heliosphere, favoring instead a {\it class of models} based on the classical developments due to \citet{vandeZiel50, Machlup81} and extended by \cite{Montroll82}. The basic idea is that an ensemble with a scale-invariant distribution of relaxation times, when superposed and sampled, gives rise to a range of $1/f$ behavior. The applicability of this reasoning is substantially extended upon the realization~\citep{Montroll82} that a log-normal distribution of relaxation times can readily produce the range of scale-invariant relaxation times required to obtain $1/f$ spectrum. We presented arguments and reviewed observations that provide evidence leading us to favor the above {\it scale-invariant superposition} explanation. We should note, once again, that our review is not exhaustive, and that statistical explanations such as {\it self-organized criticality} have been offered as a variant explanation~\citep{Bak87}.

For heliospheric $1/f$ noise, we reviewed two distinct models: a coronal model based on successive reconnections manifestly related to the superposition principle, and another model that relies on the generation of $1/f$ due to inverse cascade in the solar dynamo. The latter model may also fit the superposition class if the underlying mechanism (not reviewed here) involves successive stretch-and-fold dynamics~\citep{Vainshtein72, Rincon19}.

Both of the above models are intrinsically nonlinear and neither appears to be limited by available time for developing dynamics. This is not the case for models originating in the super-Alfv\'enic wind, where available time is limited by convection time to the position of observation. We have explained how this limited range of influence disallows inverse cascade to very long wavelengths and also limits the ability of any MHD process to access the low frequencies ($\sim \unit[10^{-6}]{Hz}$ at which $1/f$ is observed at 1 au). This is not to say that local mechanisms proposed to explain $1/f$-like spectra near $\unit[10^{-4}]{Hz}$ need to be rejected~\citep{Velli89, Chandran18, Davis23, Huang23}; it does provide a challenge to explain how the local $1/f$ spectrum matches smoothly onto the $1/f$ signal that extends to near $\unit[10^{-6}]{Hz}$. In this regard, it is useful to recall the adage from \citet{Machlup81}: {\it ``If you have not found the $1/f$ spectrum, it is because you have not waited long enough.''}

Because long data records are needed to study the $1/f$ signal, there have been relatively few in depth studies of its origins and connections to solar or coronal phenomena. The purpose of the present paper has been to assemble an overview of current knowledge about $1/f$ in the heliosphere and to point toward what we believe are the likely close connections between the solar dynamo, coronal dynamics, and effects observed in the super-Alfv\'enic wind, including 1 au and beyond. As the space physics community delves into these likely connections, the possibility may emerge that deeper knowledge of $1/f$ noise and relevant statistics might translate into quantitative connections to solar terrestrial relations and, perhaps eventually, connections to space weather prediction. At present this is a speculative remark, 
and we might even imagine that modulation of $1/f$ noise might be a trigger for rare events such as large flares or CMEs. If such connections are established, then it could represent an entirely new and statistical approach to this complex coupling between solar dynamics and geospace responses. 

Although observations and simulations are continuing to reveal aspects of $1/f$ noise in the heliosphere, it is fair to say that there remain aspects and many details that are incompletely understood. We anticipate that advanced 
instrumentation on upcoming missions such as PUNCH will provide valuable information to further reveal the mysteries of phenomena such as $1/f$ noise in the solar wind.

\begin{acknowledgments}
This research is partially supported by the NASA LWS grants 80NSSC20K0377 (subcontract 655-001) and 80NSSC22K1020, by the NASA IMAP project at UD under subcontract SUB0000317 from Princeton University, by the NASA/SWRI PUNCH subcontract N99054DS at the University of Delaware,  by the NASA HSR grant 80NSSC18K1648, and by National Science Foundation grant AGS-2108834.
\end{acknowledgments}
\newpage
\appendix
\section{$\lowercase{1/f}$ from Arbitrary Spectral Index}
\label{app:1/f}

In this appendix, we build upon the discussion in Section~\ref{subsec:Machlup} to demonstrate that the superposition-based generation mechanism for $1/f$ noise is applicable to a time series ensemble with any index $-\alpha$, where $\alpha > 1$. A straightforward approach~\citep[][]{Matthaeus86} is to assume, in place of the Lorentzian profile in Eq.~\ref{eq:Stau}, the following form for individual time series' spectrum:
\begin{equation}
    S(\tau, \omega) \propto \frac{\tau}{(1+\omega^2 \tau^2)^{\alpha/2}},
\label{eq:Salpha2}
\end{equation}
where $S(\omega)$ is flat at small frequencies ($\omega \tau \ll 1$) and has a power-law index of $-\alpha$ at large frequencies ($\omega \tau \gg 1$). With inversely distributed correlation times as in Eq.~\ref{eq:inverse}, the superposed spectrum becomes
\begin{equation}
    \overline{S}(\omega) \propto \frac{1}{\omega} \int_{x_1}^{x_2} \frac{dx}{(1+x^2)^{\alpha/2}} = \frac{1}{\omega} \left[ \int_{0}^{\infty} \frac{dx}{(1+x^2)^{\alpha/2}} - \int_{0}^{x_1} \frac{dx}{(1+x^2)^{\alpha/2}} - \int_{x_2}^\infty \frac{dx}{(1+x^2)^{\alpha/2}} \right],
\label{eq:Soverline2}
\end{equation}
where $x \equiv \omega \tau$, and in the frequency range of interest, $x_1 \equiv \omega \tau_1 \ll 1$ and  $x_2 \equiv \omega \tau_2 \gg 1$. The first integral in Eq.~\ref{eq:Soverline2} has a definite solution of $\sqrt{\pi} \Gamma[(\alpha-1)/2]/2 \Gamma(\alpha/2)$ where \(\Gamma\) denotes the gamma function. The third integral can be approximated as $\int_{x_2}^\infty dx/x^\alpha = x_2^{1-\alpha}/(\alpha-1)$ for $x_2 \gg 1$. The solution to the second integral is proportional to the hypergeometric function, $x_1 {}_2F_1 (1/2, \alpha/2; 3/2; -x_1^2 ) = x_1 - \alpha x_1^3 / 6 + \mathcal{O}(x_1^5)$. Thus 
\begin{equation}
    \overline{S}(\omega) \propto \frac{1}{\omega} \left[ \frac{\sqrt{\pi} \Gamma (\frac{\alpha-1}{2})}{2 \Gamma(\frac{\alpha}{2})} - \mathcal{O}(x_2^{1-\alpha}) - x_1 + \mathcal{O}(x_1^3) \right].
\end{equation}
To the least order approximation, $\overline{S}(\omega) \propto 1/\omega$ only if $\alpha > 1$, that is, the order of the second term in the bracket is less than that of the first term. The constraint of $\alpha > 1$ also ensures that $\Gamma[(\alpha-1)/2]$ is positive. 

We now propose an alternative derivation of $1/f$ noise that avoids the hypergeometric function (as well as other complicated math) and avoids assuming a certain spectral form, as in Eq.~\ref{eq:Salpha2}. Instead, we assume that the spectrum is flat below a certain break frequency and transitions to a power-law at higher frequencies. The main idea is to evaluate the expected slope of $\log(\overline{S})(\log(\omega))$ through assigning a weighting function for the slopes at any given $\omega$. To maintain clarity in notation, we now use $\tau_c$ to represent the correlation time for an individual time series. The associated frequency, at which the power spectrum transitions from a flat profile to a power-law decay with an index of $-\alpha$, is denoted as $\omega_c \equiv 1/\tau_c$. Meanwhile, $\omega$ indicates the range of frequencies within the domain of interest.

The reciprocal of an inversely distributed random variable is also inversely distributed. Indeed $\omega_c$ follows the distribution 
\begin{equation}
    \rho(\omega_c) d\omega_c = \frac{d \omega_c}{\omega_c \log(\omega_1/\omega_2)},
\end{equation}
where $\omega_1 \equiv 1/\tau_1 > \omega_2 \equiv 1/\tau_2$, and $\omega_c \in [\omega_2, \omega_1]$. For each power density spectrum normalized to an equal total power (of unity for example), the height of the power spectral density at $\omega_c$ is denoted as $S_c$ and is constant at lower frequencies. The values of $S_c$ are inversely proportional to $\omega_c$, and are inversely distributed as
\begin{equation}
    \rho(S_c) dS_c = \frac{dS_c}{S_c \log{(\omega_1/\omega_2)}}.
\end{equation}

Above the break frequency, a power spectrum follows the form $S(\omega) = S_c(\omega/\omega_c)^{-\alpha}$ assuming continuity. Therefore, at any given frequency $\omega$, the power-law index, denoted as $\beta$, is either $0$ if the chosen spectrum has $\omega_c > \omega$ or $-\alpha$ if $\omega_c < \omega$. Intuitively, the expected value of $\beta$ is
\begin{equation}
    \overline{\beta} (\omega) = \frac{\int_{S_c(\omega_1)}^{S_c(\omega_2)} \beta(\omega) S(\omega) \rho(S_c) dS_c }{ \int_{S_c(\omega_1)}^{S_c(\omega_2)} S(\omega) \rho(S_c) dS_c  },
\label{eq:beta}
\end{equation}
where $S_c(\omega)$ represents the magnitude $S_c$ given the spectrum has a break frequency at $\omega$. Here $\rho(S_c)$ can be considered as the probability density of $S(\omega)$, and $S(\omega)$ is the weighting function of $\beta$ at frequency $\omega$. 

Why is $S(\omega)$ the weighting function of $\beta (\omega)$ under the context of spectrum superposition? To show this, consider two spectra $S_\mu (\omega) \propto (\omega/\omega_\mu)^\mu$ and $S_\nu (\omega) \propto (\omega/\omega_\nu)^\nu$ of power-law indices $\mu$ and $\nu$, respectively, where $\omega_\mu$ and $\omega_\nu$ are arbitrary positive constants. The slope of $S_\mu + S_\nu$ in log-log scale can be directly calculated as
\begin{equation}
    \frac{\partial}{\partial{\log{(\omega)}}} \log{(S_\mu + S_\nu)} = \frac{d \omega}{d \log{(\omega)}} \frac{\partial}{\partial{\omega}} \log{(S_\mu + S_\nu)} = \frac{\mu S_\mu + \nu S_\nu}{S_\mu + S_\nu}.
\end{equation}
We have shown that on log-log scale, power indices of power spectra can be computed with a weighted average.

The expectation $\overline{\beta} (\omega)$ can now be confidently computed using Eq.~\ref{eq:beta}, keeping in mind that $\beta = -\alpha$ when $S_c>S(\omega)$ and $\beta = 0$ otherwise:
\begin{equation}
    \overline{\beta} (\omega) = \frac{-\alpha \int_{S_c(\omega)}^{S_c(\omega_2)} S_c (\omega/\omega_c)^{-\alpha} \rho(S_c) dS_c }{ \int_{S_c(\omega_1)}^{S_c(\omega)} S_c \rho(S_c) dS_c + \int_{S_c(\omega)}^{S_c(\omega_2)} S_c (\omega/\omega_c)^{-\alpha} \rho(S_c) dS_c } = -\alpha \left[ 1 + (1-\alpha) \frac{1-(\omega/\omega_1)}{(\omega_2/\omega)^{\alpha-1}-1}  \right]^{-1}.
\end{equation}
Within the frequency region where $\omega_2 \ll \omega \ll \omega_1$, and assuming $\alpha > 1$, we arrive at the desired result of $\overline{\beta} (\omega) = -1$.


\end{CJK*}
\end{document}